\documentclass[aps,prl,reprint,showpacs,superscriptaddress]{revtex4-1}
\usepackage{graphicx,color}
\usepackage{amssymb,amsmath}
\usepackage[T2A]{fontenc}
\usepackage[cp1251]{inputenc}

\begin{document}
\title{The impact of quantum effects on relativistic electron motion in a chaotic regime}

\author{A.V. Bashinov}
\affiliation{Institute of Applied Physics, Russian Academy of Sciences, 603950 Nizhny Novgorod, Russia}
\affiliation{University of Nizhny Novgorod, 603950 Nizhny Novgorod, Russia}
\author{A.V. Kim}
\affiliation{Institute of Applied Physics, Russian Academy of Sciences, 603950 Nizhny Novgorod, Russia}
\affiliation{University of Nizhny Novgorod, 603950 Nizhny Novgorod, Russia}
\author{A.M. Sergeev}
\affiliation{Institute of Applied Physics, Russian Academy of Sciences, 603950 Nizhny Novgorod, Russia}
\affiliation{University of Nizhny Novgorod, 603950 Nizhny Novgorod, Russia}
\date{\today}

\begin{abstract}
We consider the impact of quantum effects on electron dynamics in a plane linearly polarized standing wave with relativistic amplitudes. For this purpose
analysis the Lyapunov characteristic exponent spectrum with and without allowance for the classic radiation reaction force has been analyzed. Based on this analysis it is concluded that the contraction effect of phase space in the stochastic regime due to the radiation reaction force in the classical form doesn't occur when the quantum nature of hard photon emission is taken into account. It is shown that electron bunch kinetics has a diffusion solution rather than the d'Alambert type solution as in the classic description.It is also revealed that the electron motion can be described using the Markov chain formalism. This method gives exact characteristics of electron bunch evolution, such as motion of the center of mass and electron bunch dimensions.
\end{abstract}

\maketitle

\section{Introduction}
Nowadays, much attention is paid to the interaction of relativistically strong laser fields with matter, when radiation losses play an important role. The interest in this subject is connected not only with the fundamental problem of charged particle dynamics in electromagnetic fields, but also with possible applications, such as laser-based electron accelerators and hard photon sources \cite{Hooker_NP,Phuoc_NP}. It was recently found that the dynamics of charged particles in ultrarelativistic electromagnetic fields changes drastically when radiative reaction force is taken into account. As was shown in Refs.\cite{RT_Pukhov,ART_Gonoskov}, at super strong laser fields charged particles are drawn into a high-field region, thus leading to efficient generation of gamma rays and particle beams. The need to consider the back impact of the emitted photon !fields! is dictated by the facts that the recoil force becomes comparable with the Lorentz force and a particle can lose a substantial part of its energy in one act of emission. Thus, the effect of radiation losses can dramatically change the trajectory of the particle over the laser field period. In some cases, small changes of particle trajectories over the laser period caused by weak radiative losses can be accumulated and significantly change particle motion for a time much longer than the optical cycle. This follows from the theorem of phase space contraction in the presence of dissipative forces, that has been proved in Ref.\cite{Tamburini_NuclInst} for the case of charged particle motion in an electromagnetic field, with the classical radiation reaction force taken into consideration. Particular examples of this effect were considered in the field of traveling \cite{QE_Bashinov} and standing waves \cite{Lehmann_PRE}. In the latter case, it is connected with stochastic heating \cite{Sheng_PRL,Medonca_JOP,Sentocu_APB}. However, due to the quantum nature, emission occurs randomly in the form of discrete photons that may randomize motion. As a consequence, the phase volume may be not compressed and even increases in spite of dissipation. Comparative analysis of an electron beam interaction with a counterpropagting laser pulse, taking into consideration radiative losses in classical and quantum form showed that at certain electron energies and laser field intensity, the phase space volume can be expanded during the interaction \cite{Neitz_PRL}. 
In this paper, we analyze the characteristics of electron motion in the field of a standing linearly polarized plane wave taking into account radiative losses in conformity with the quantum and classical descriptions. We consider the wave amplitudes at which an electron loses a small part of its energy in one act of photon emission but its motion can qualitatively change due to discreteness of the photon emission. A comparative analysis of electron dynamics with and without radiation reaction force in classical and quantum forms is presented.

\section{Electron motion in the classical case} 

For taking into account radiative losses in the classical case, we add to the equation of motion the radiation reaction force in the Landau-Lifshitz form, keeping the most important term proportional to the squared Lorentz factor \cite{Tamburini_NJP}:
\begin{gather}
 \frac{d\bf p}{dt}=-e{\bf E}-\frac{e}{c}{\bf v\times B}-\nonumber\\ 
\label{landlifp}\frac{\delta e^2\gamma^2 {\bf v}}{\omega mc^2}\left[\left({\bf E}+\frac{1}{c}{\bf v \times B}\right)^2-\frac{1}{c^2}\left({\bf E\cdot v}\right)^2\right],\\
\label{landlifr} \frac{d\bf r}{dt}=\frac{\bf p}{m\gamma},
\end{gather}
where $p,~v,~\gamma,~-e,~m$ are the momentum, velocity, Lorentz factor, charge and mass of the electron, correspondingly;  $c$ is the velocity of light; $t$ is the time in the laboratory frame of reference; $\delta=2e^2\omega/(3mc^3)$; and $\omega$ is the laser frequency. Without loss of generality, we assume that the electric field is directed along the Z axis and the magnetic field is directed along the X axis:
\begin{gather}
\label{ef} {\bf E} = E_0\cos (\omega t+\varphi_0) \cos(ky) {\bf z},\\
\label{bf} {\bf B} = E_0\sin (\omega t+\varphi_0) \sin(ky) {\bf x},
\end{gather}
where the vacuum wave number $k=\omega/c$ and $\varphi_0$ is the initial wave phase. The electron density is considered to be low enough, so that the effect of the particles influence on each other can be neglected in comparison with the impact of the external field. Dimensionless variables $\eta=ky,~\zeta=kz,~\rho=p/mc,~\tau=\omega t,~\upsilon=v/c,~a_0=eE_0/(m\omega c)$ are used for simplicity; we also assume that at the initial moment of time, the particle moves in the ZY plane. In this case, its motion always occurs in this plane:
\begin{gather}
\label{llpy} \gamma\frac{d\rho_y}{d\tau}=-\rho_za_0\sin (\tau+\varphi_0) \sin(\eta)-\delta\rho_ya_0^2\Theta(\tau,\eta),\\
 \gamma\frac{d\rho_z}{d\tau}=\rho_ya_0\sin (\tau+\varphi_0) \sin(\eta)-\nonumber\\
\label{llpz}\gamma a_0\cos(\tau+\varphi_0)\cos(\eta)-\delta\rho_za_0^2\Theta(\tau,\eta),\\
\label{lly} \frac{d\eta}{d\tau}=\rho_y/\gamma,\\
\label{llz} \frac{d\zeta}{d\tau}=\rho_z/\gamma,
\end{gather}
where $\Theta=(\gamma^2-\rho_z^2)\cos^2(\tau+\varphi_0)\cos^2(\eta)+\rho^2\sin^2(\tau+\varphi_0)\sin^2(\eta)-\frac{\rho_y}{2}\sin(2(\tau+\varphi_0))\sin(2\eta)$, $\rho=\sqrt{1+\rho_y^2+\rho_z^2}$, $\gamma=\sqrt{1+\rho^2}$. To analyze the electron motion without radiation reaction force it suffices to assume $\Theta=0$. Note that, although further analysis is performed for the wavelength $\lambda=0.8 \mu m$, when $a\gg1$, the system depends to a good accuracy only on one similarity parameter $\delta a^3=\frac{4\pi r_e}{3\lambda}a^3$ \cite{Bulanov_PRL}.\par
\begin{figure}[h]
	\centering
		\includegraphics[width = 0.23\textwidth]{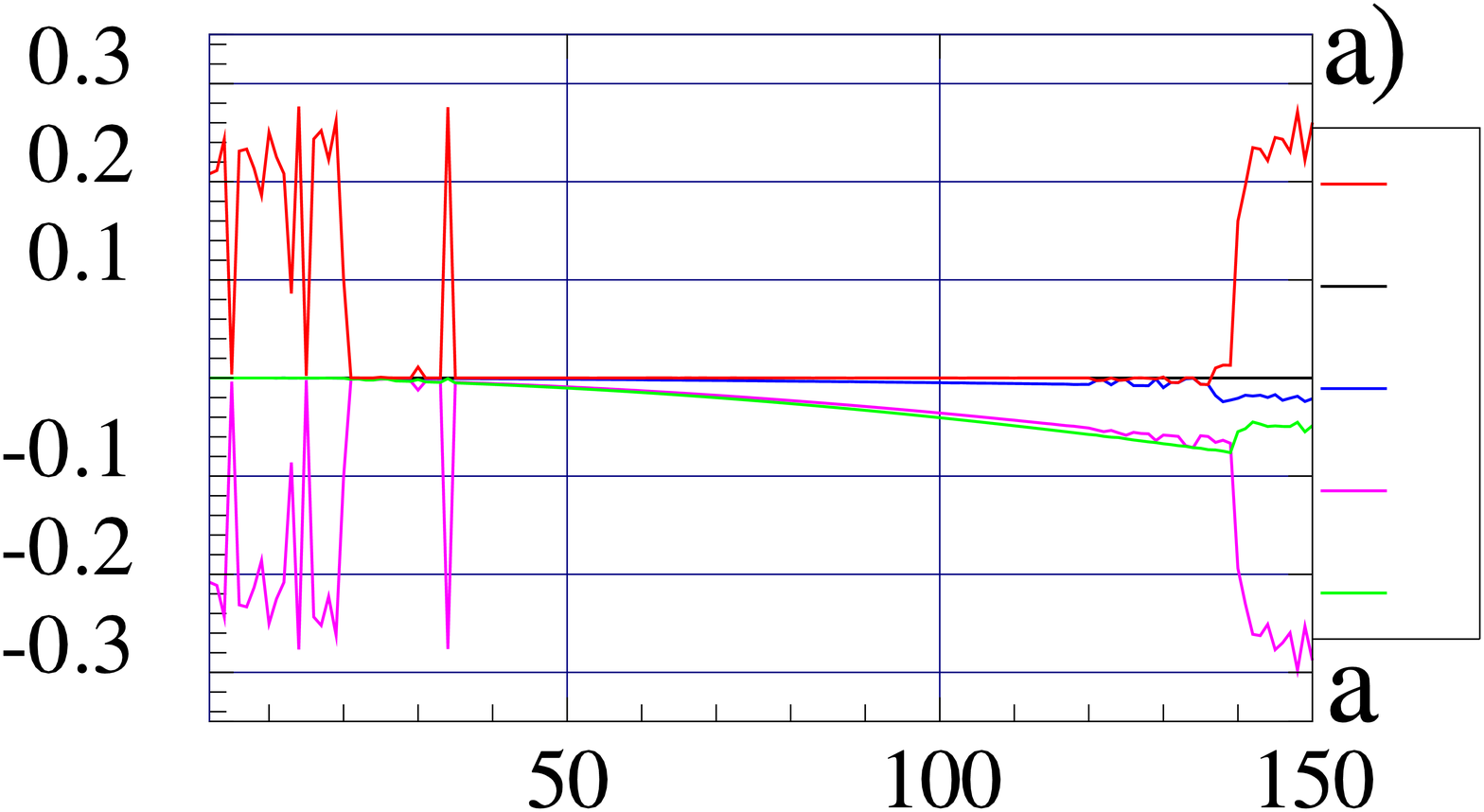}
		\includegraphics[width = 0.23\textwidth]{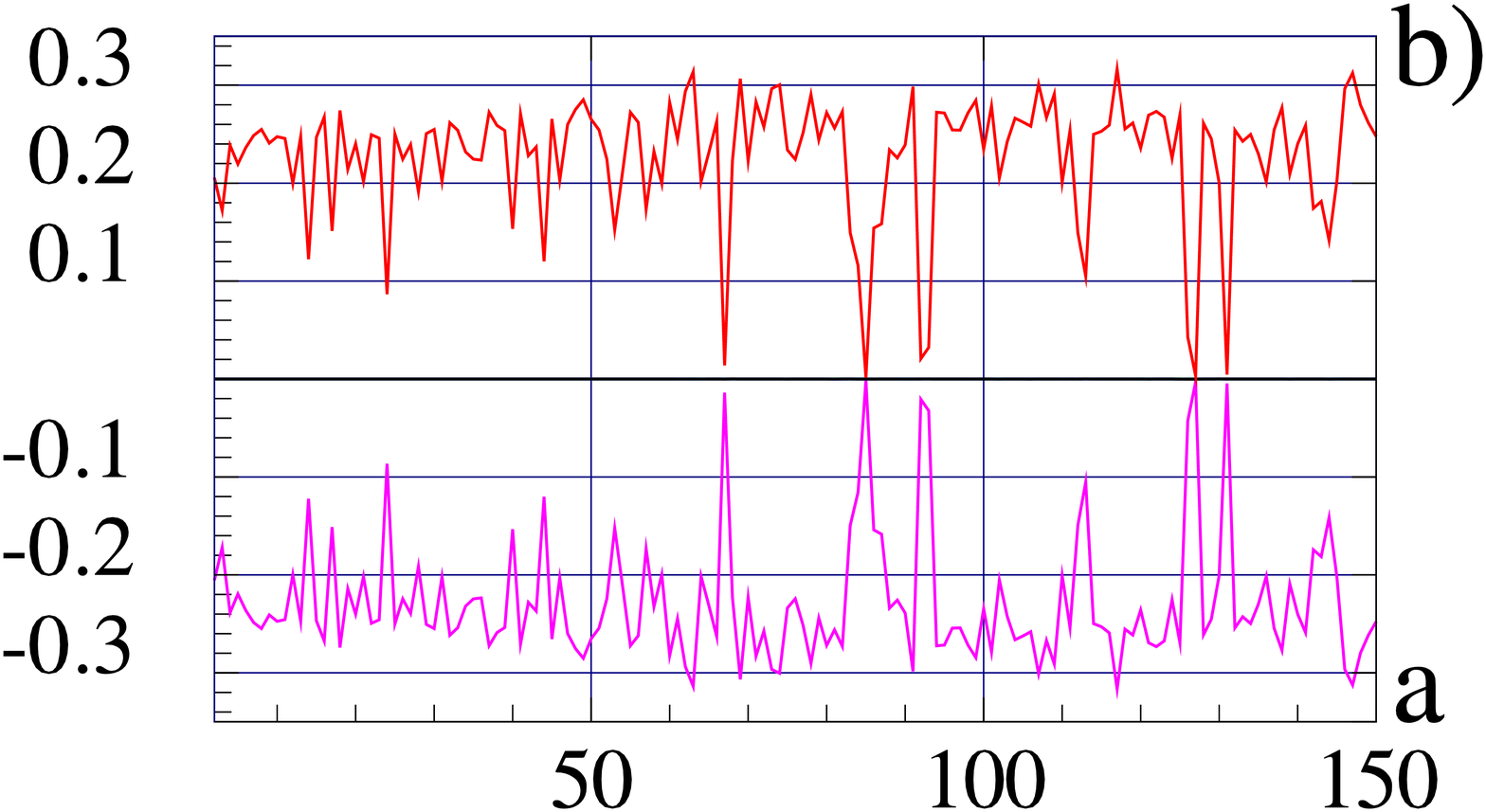}
	\caption{Lyapunov characteristic exponents of system (\ref{llpy}-\ref{lly}) with (a) and without ($\Theta=0$) (b) radiation reaction force. According to value/По старшинству $\lambda_1$ (red), $\lambda_2$ (black), $\lambda_3$ (blue), $\lambda_4$ (magenta). The sum of the exponents is $\lambda_g$ (green).}
		\label{specLyap}
\end{figure}

In a general case,  analytical solution of the equation of motion (\ref{llpy}-\ref{llz}) in the field of a plane linearly polarized standing wave hasn't been found, even without radiation reaction, so the spectrum of characteristic Lyapunov exponents is used to analyze electron motion with the help of a modified generalized Benettin algorithm \cite{Benettin_1},\cite{Benettin_2}. In order to find all possible modes of motion, it is necessary to consider all possible initial conditions. Because of the periodicity of the equations along the Y axis it is sufficient to set the initial position of the electrons along the axis within one wavelength. Taking into account the radiative reaction force, it is reasonable to assume that the absolute value of the transverse momentum (along $\bf{E}$) during steady motion doesn't exceed $2a$, and the absolute value of the longitudinal momentum doesn't exceed  $a^2/2$. However, significant computing resources are needed for good resolution of phase space by initial conditions. We confine ourselves to a single particle, which has zero momentum at the initial moment of time and is located between the electric and magnetic field nodes (in the calculations, the particle was set midway between the nodes). Thus, we haven't obtained all possible modes of motion (for example, stable trajectory in the region of the electric field antinode at a relativistic initial momentum along the electric field \cite{Kaplan_PRL}), but such initail position and momentum allow the particle to sit on attractors, which are discussed in the work \cite{Lehmann_PRE}.
To reduce the influence of the initial conditions, the calculation of the Lyapunov characteristic exponents starts only after 30000 periods of laser field. As shown by calculations, this time is sufficient for the particle trajectory to sit on the attractor, if it exists, even with $a\sim1$. Figure \ref{specLyap} shows the obtained spectra of Lyapunov characteristic exponents as a function of $a$.
First, note that, as the dimension of system (\ref{llpy}-\ref{llz}) is 4, there exist chaotic attractors. Second, the results of numerical modeling are consistent with the theorem of phase space contraction \cite{Tamburini_NuclInst}, taking into account the classic radiation reaction: the sum of the Lyapunov exponents is negative, despite the fact that the maximum Lyapunov exponent can be positive, the Lyapunov exponents equal zero. Consequently, system (\ref{llpy}-\ref{llz}) has a chaotic or a regular attractor. Note that there is a range of amplitudes $35\le a<138$ where the attractor is regular, which is conformed by the conclusion in \cite{Lehmann_PRE}. The stochastic heating is suppressed in this amplitude range. Since the sum of Lyapunov exponents is negative, one of them is assumed to be zero, which improves accuracy of estimations of the other Lyapunov exponents. Average squared error of the exponents in this case did not exceed 1$\%$.\par
But without the radiation reaction, system (\ref{llpy}-\ref{lly}) may have attractors under certain initial conditions \cite{Sheng_PRL}, \cite{Kaplan_PRL}. However, the calculations havn't given us definite information at what amplitudes attractors may arise. The exact value of the sum of the exponents $\lambda_g$ could not be established because $\lambda_g\sim10^{-9}\pm0.01$, i.e. the error is much greater than the value itself. This occurs as the largest and the smallest absolute values of the Lyapunov exponents are opposite in sign $|\lambda_{1,4}|\sim0.2\pm0.002$, and the second and third parameters are close to zero $\lambda_{2,3}\sim10^{-6}\pm10^{-4}$. However, the value of the sum of the Lyapunov exponents isn't of great significance, most important is that the senior indicator is positive and there is chaotic motion without radiation friction \cite{Lehmann_PRE}.\par
Note that the spectrum of the characteristic Lyapunov exponents is calculated for a certain trajectory that originates under certain initial conditions. However, according to the multiplicative ergodic theorem \cite{Oseledec_68}, this spectrum can be attributed to the attractor as a whole. We have considered the attarctor which electron initially at rest reaches. Moreover if the initial momentum is  $p_0\ll a$, the particle goes to the same attractor \cite{Lehmann_PRE}. We also note that access to the attractor, as follows from numerical calculations, does not depend on the initial phase of the wave. The initial phase affects the steady component of the pulse along the electric field ($p_z$), but due to radiation losses the steady component disappears in some time that will be estimated later. As a result, the average value of $p_z$ is equal to 0, and that's why the initial phase has no influence on a possibility of reaching the attractor. However, there are points in phase space starting at which an electron cann't reach the attractor. The magnetic field antinodes $y_B=0.25\lambda+0.5n\lambda$ (where n is an integer) are stable at small perturbations. In the nonrelativistic case, the electron motion near this point can be described as damped oscillations in the ponderomotive potential due to radiation reaction force. It can be assessed in dimensionless units that $|p|\ll1$ along with $|y-y_B|\ll \pi/2$ and $a>1$ are sufficient for stability. The momentum can be assessed as $p_y\approx p_z\le |p_{0}|+2a|y-y_B|$, since a trajectory is rotated in a magnetic field with consecutive increasing and decreasing of energy due to the electric field work.
Therefore, a particle with zero initial momentum is captured by the antinode of the magnetic field, if $|y_0-y_B|<1/2a$. Due to the small size of this area around the antinode of the magnetic field, from which the particle will not go away, only a small fraction of the uniformly distributed particles is captured by this area. There is also an unstable point of equilibrium $y_E=0.5n\lambda$, however, the arbitrarily small perturbations allow the particle to reach the attractor. Thus, only a small amount of particles does not reach the found attractor. In the case without radiation reaction, the initial phase of the standing wave comes into play. For comparison of the results of the calculation of Lyapunov exponents with and without radiation reaction, the initial phase is chosen to be 0, which corresponds to a zero steady component of the transversal momentum.
Without radiation reaction, the antinodes of the electric field are also unstable. The magnetic field antinodes are in indifferent equilibrium. Starting with an arbitrary position outside a small region around the magnetic field antinode, the trajectories are qualitatively similar (for a long period of time, the trajectories fill the same space of phase projection $p_yp_zy$). So, the Lyapunov exponent spectrum of a trajectory can be generalized to the others.\par
\begin{figure}[h]
	\centering
		\includegraphics[width = 0.23\textwidth]{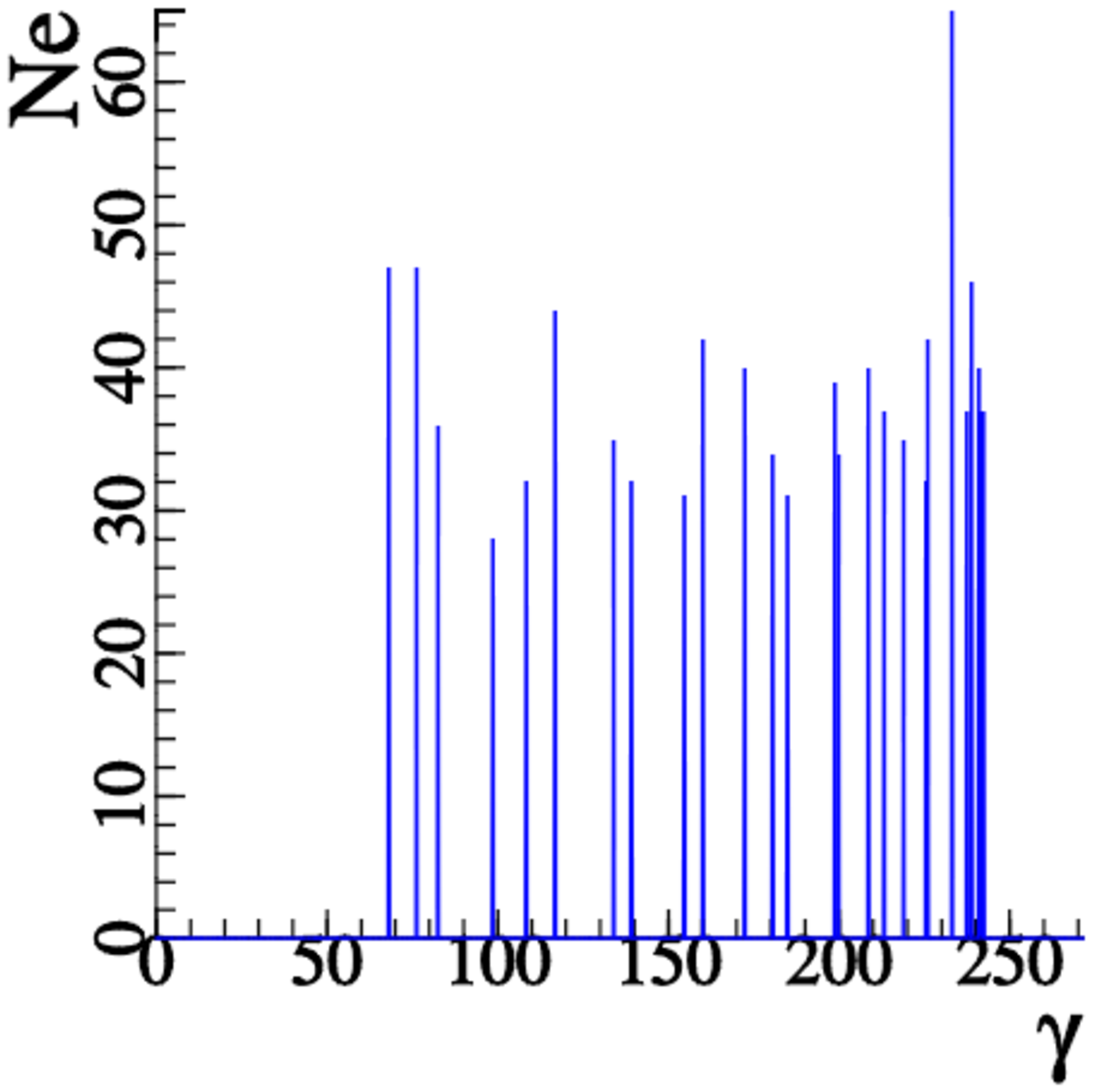}
		\includegraphics[width = 0.23\textwidth]{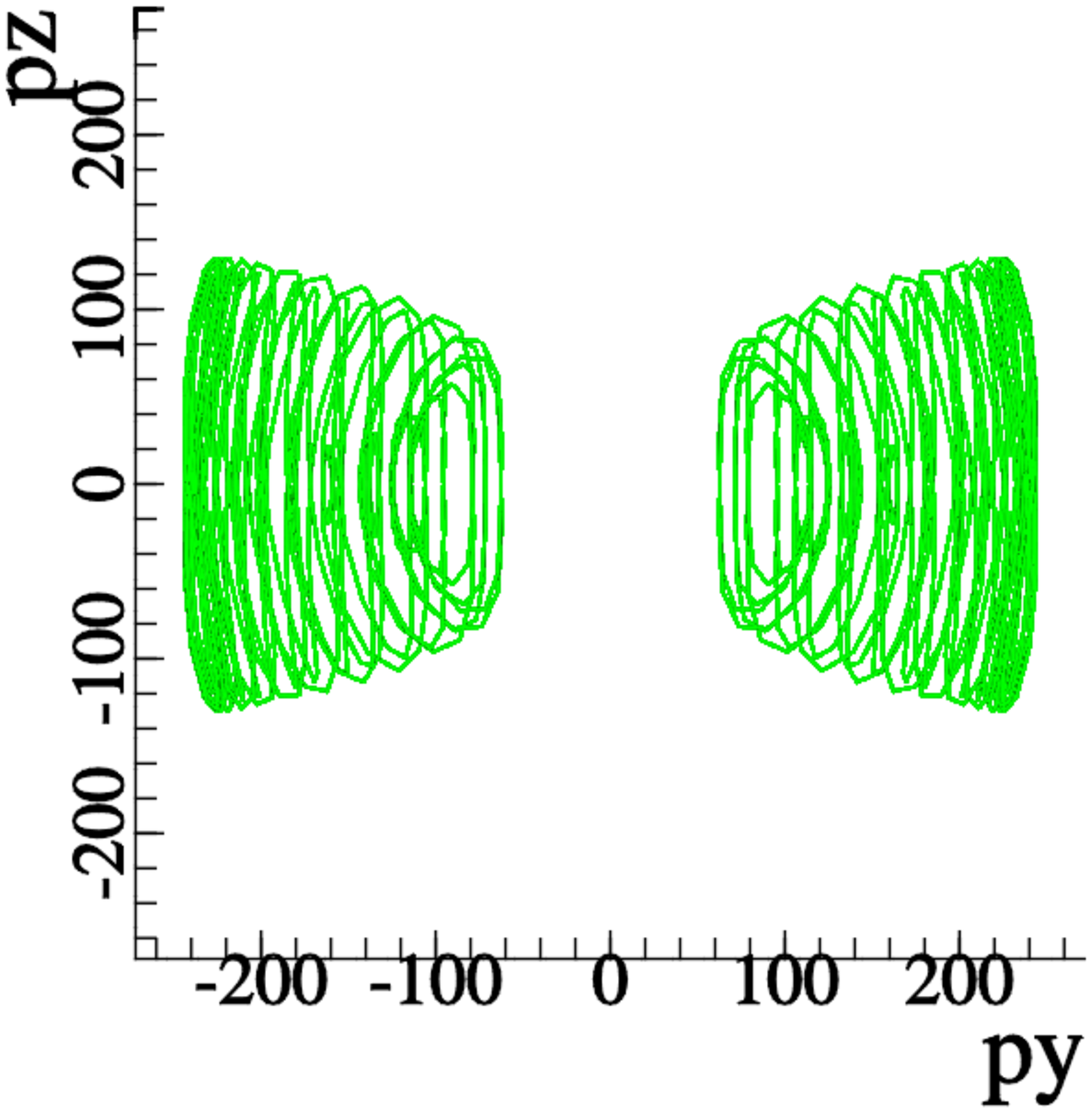}
	\caption{Elelctron energy distribution in the field of a plane wave with amplitude $a=136$ (a), and phase space projection on the $p_yp_z$ plane (b) with classical description of radiation.}
	
	\label{fig_af_68_50En}
\end{figure}
\begin{figure}[h]
	\centering
		\includegraphics[width = 0.23\textwidth, height =0.15\textwidth]{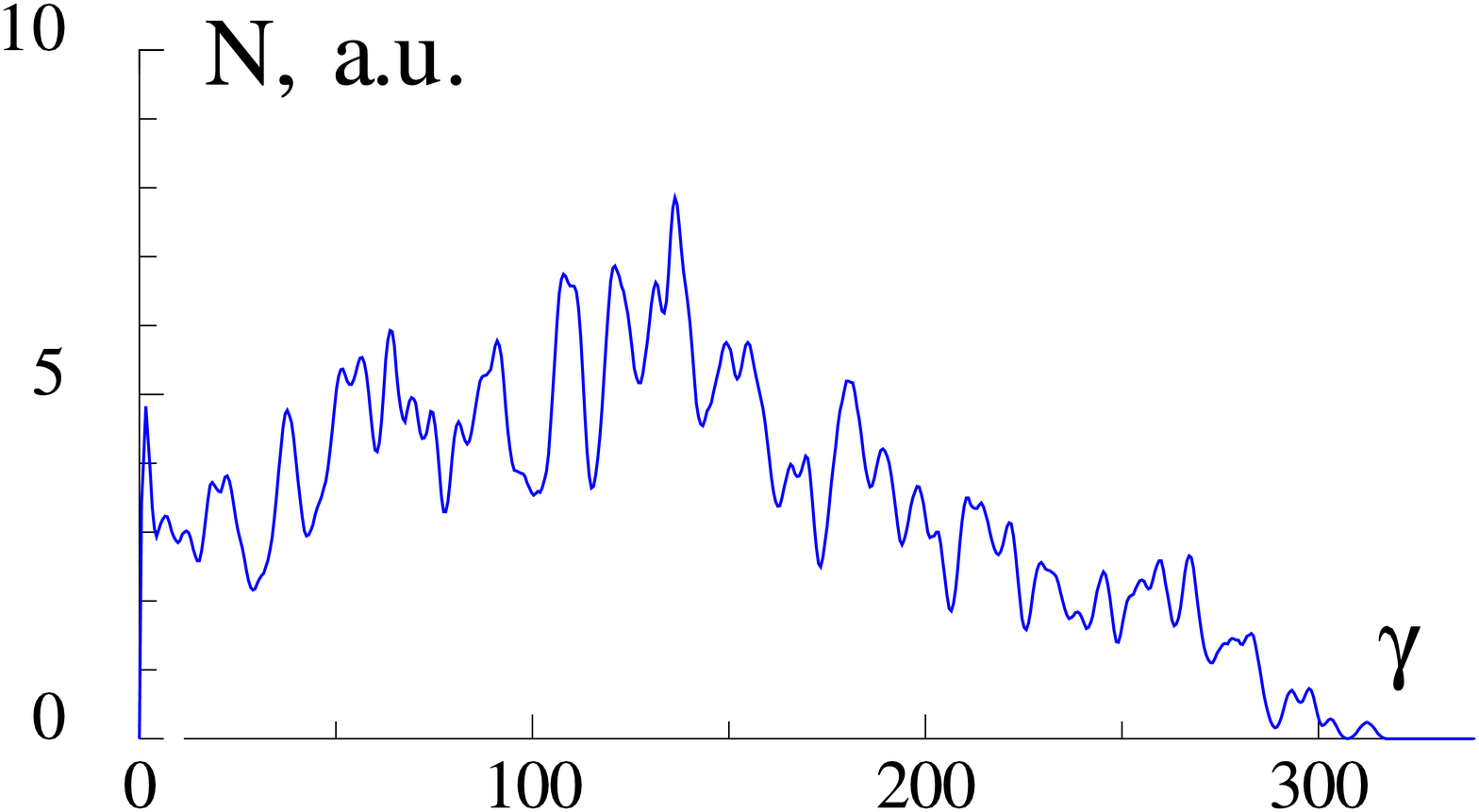}
		\includegraphics[width = 0.23\textwidth, height = 0.15\textwidth]{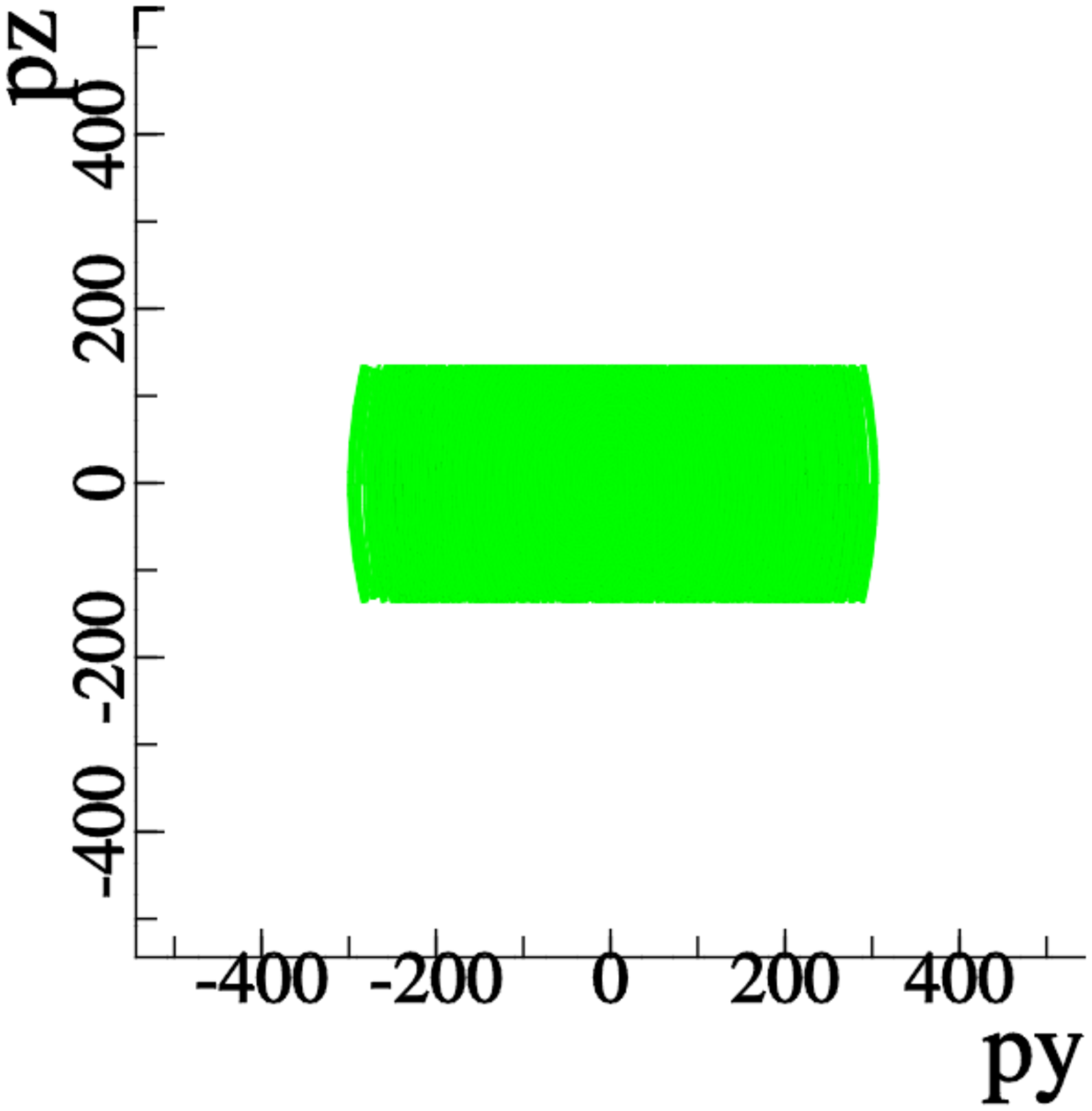}	
	\caption{Electron energy distribution in the field of a plane wave with amplitude $a=136$ (a), and phase space projection on the $p_yp_z$ plane (b) without radiation reaction.}
	
	\label{fig_af_65_50Enwo}
\end{figure}
\begin{figure}[h]
	\centering
		\includegraphics[width = 0.45\textwidth, height=0.2\textwidth]{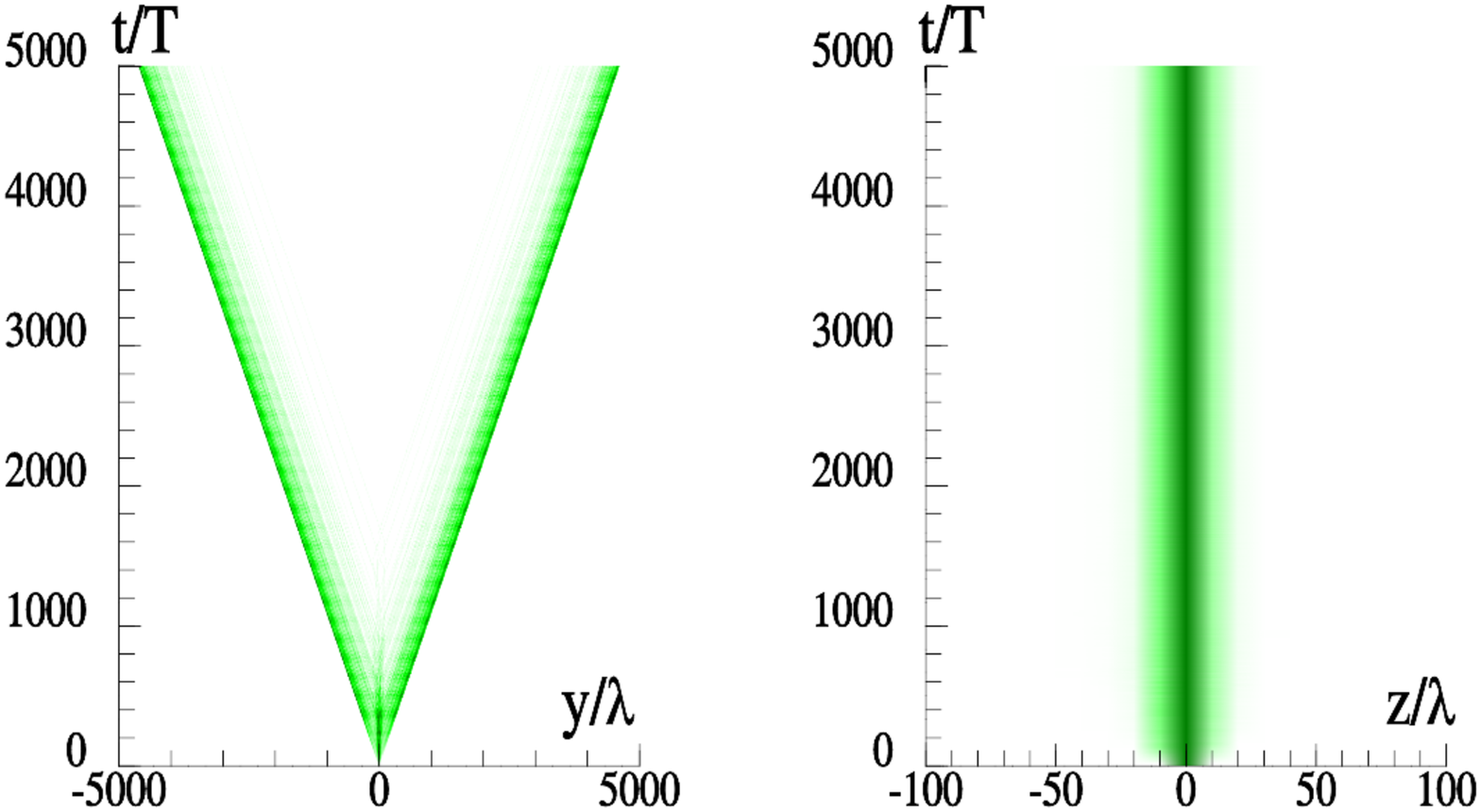}
		\includegraphics[width = 0.45\textwidth, height=0.2\textwidth]{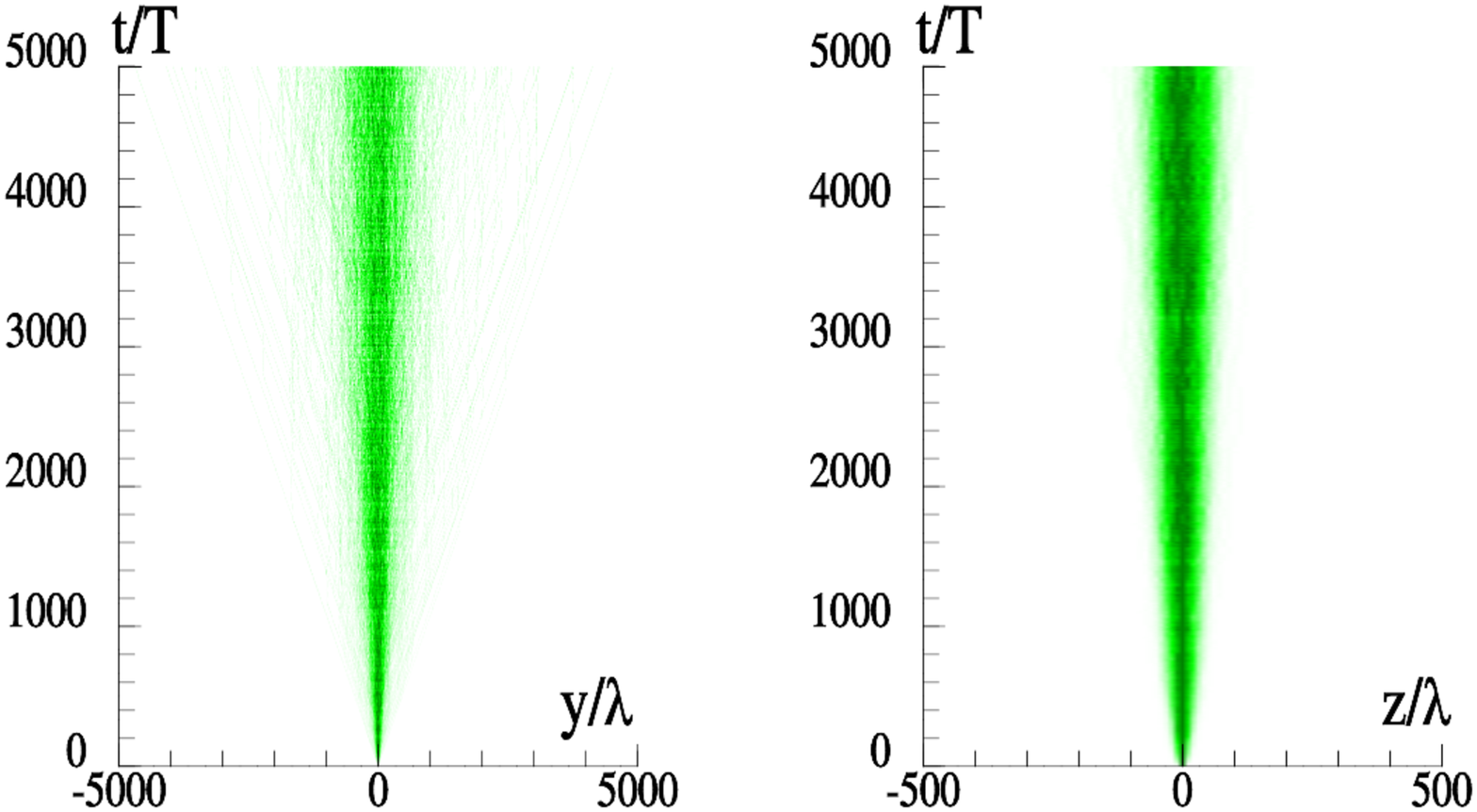}
	
	\caption{Time dependence of electron distribution along Y and Z axes for $a=136$ with (a) and without (b) radiation reaction force.}
	\label{136YZt}
\end{figure}

Note that each characteristic parameter $\lambda_i$ corresponds to the basis vector $\vec{i}$ and along these vectors the trajectory perturbation increases or decreases as $e^{\lambda_i t}$, depending on the sign of $\lambda_i$. According to numerical simulation, the same basis vector (\ref{rb}) corresponds to the highest nonzero Lyapunov exponent with and without radiation reaction. The vector $r_b$ is determined to an accuracy of 5$\%$ in the range of standing plane wave amplitudes $35\le a<138$. Disturbances along this vector decrease most slowly (the Lyapunov exponent is negative, the radiation reaction is taken into account) or increase most rapidly (the Lyapunov exponent is positive, the radiation reaction is neglected):\par
\begin{equation}
 \vec{r_{b}}=(1i_{p_y},0i_{p_z},0i_{t},0i_{y})
 \label{rb}
 \end{equation}
Compare the dynamics of electrons in the field of a standing plane wave with and without radiation reaction. We consider the wave amplitudes at which a regular attractor is possible assuming the radiation reaction to be taken into account. If the attractor is chaotic, no significant difference is observed in the motion of an electron with and without radiation reaction force. Moreover, the maximum Lyapunov exponents are approximately the same in these two cases.
\begin{figure}
	\centering
		\includegraphics[width = 0.49\textwidth]{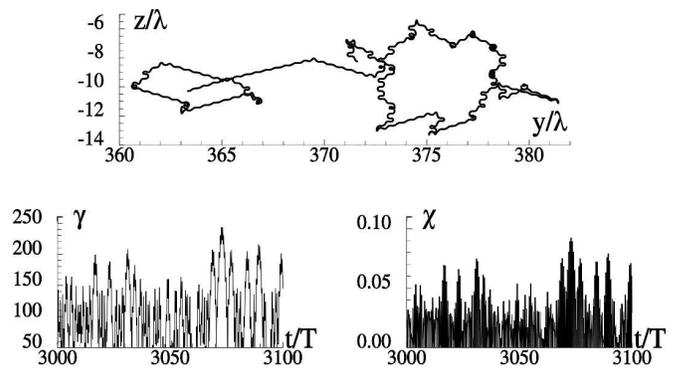}
	
	\caption{Trajectory of an electron in the field of a linearly polarized standing wave with amplitude $a=136$ initially placed between electric and magnetic field nodes with zero momentum after 3000T and $\gamma$ and $\chi$ as a function of time without radiation losses.}

	\label{fig:traj0}
\end{figure}
\begin{figure}
	\centering
		\includegraphics[width = 0.5\textwidth]{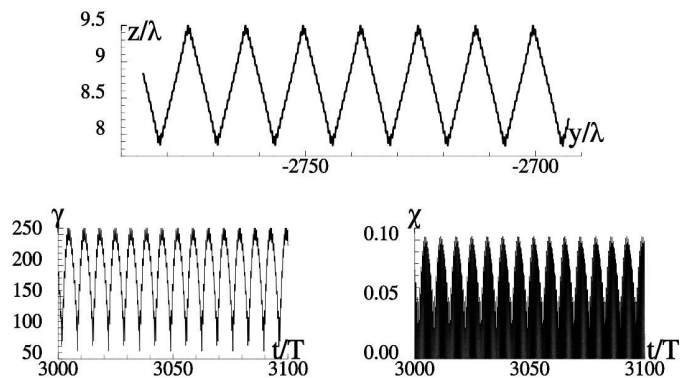}
	\caption{Trajectory of an electron like in Fig.\ref{fig:traj0} but with classical radiation reaction force.}
	
	\label{fig:traj1}
\end{figure}
For numerical analysis it is convenient to consider the wave amplitude corresponding to the shortest time needed for the particle to approach the attractor; it occurs when $a=136$. In this case, the largest non-zero Lyapunov exponent $\lambda_2=-0.0067$. Figure \ref{fig:traj0},\ref{fig:traj1} shows the trajectories of the electron in the field of a plane wave with amplitude $a=136$, with and without radiation reaction taken into account, respectively, after 3000 field periods. This time is sufficient for the particle motion to converge to the attractor. The particle motion along the attractor is directed, the averaged transversal momentum is zero (there is no drift in the transversal direction), and the longitudinal momentum oscillates around $\sim150mc$ with an amplitude of $\sim100mc$ and an averaged longitudinal velocity of $0.92c$. Note that the averaged value of energy is $<\gamma>\sim amc^2$. Without radiation reaction, the electron motion is partially similar to the electron motion with radiation reaction taken into account, but there is a possibility for the electron to be trapped by a magnetic field antinode. When trapped, it stays in this state for random time, drifting along the electric field, thus making the motion chaotic. This is the reason of directed motion loss. So, if we consider the electron bunch motion in the field of a standing wave with allowance for the radiation reaction, then we will have a "wave-like solution"' (the electrons are halved and move in opposite directions along the Y axis). In this case, after a while there are no electrons in the initial region. However, without radiation reaction, electron bunch spreading is observed (a "diffusion like solution") due to the random change of electron motion direction. Let us analyze the electron bunch motion. 1000 electrons are initially uniformly distributed within the region $-0.25\lambda<y<0.25\lambda$ and $-0.1\lambda<z<0.1\lambda$. The initial momentum is zero. Figure \ref{fig_af_68_50En}-\ref{136YZt} shows the results of calculations. The quantum parameter $\chi\ll1$ \cite{Ritus_85},\cite{Nikishov_85}, i.e. the classical approach is applicable for description of the electron motion:\par
\begin{equation}
\label{chi_exact}
\chi = \frac{e\hbar}{m^3c^4} \sqrt{\left(mc\gamma\textbf{E} + \textbf{p}\times\textbf{H}\right)^2 - \left(\textbf{p}\cdot\textbf{E}\right)^2}
\end{equation}
The energy spectrum of the electrons, with the radiation reaction taken into account, is discrete, and the spectrum line thickness is less than 0.2MeV. The number of spectrum lines is determined by the time needed for the electron to approach the attractor, as well as by the fact that the electron motion is quasiperiodic and its period equals several periods of the standing wave ($\sim14$ in the case considered). The average energy of the electrons along the attractor is $\sim$80MeV. The projection of the phase space portrait on the $p_yp_z$ plane is a line that is the result of contraction of the phase space in Fig.\ref{fig_af_68_50En}. In particular, this figure confirms that the major part of particles initially uniformly distributed within the $-0.25\lambda<y<0.25\lambda$ range converge to the same attractor. In other words, for the majority of particles their initial position is of no significance, which justifies our choice of the initial conditions for estimating Lyapunov exponents. The phase portrait without radiation reaction is "blurred", the spectrum becomes continuous and shifted to lower energies due to a random change of electron motion direction, and the average energy decreases to 60 MeV (Fig. \ref{fig_af_65_50Enwo}). In the absence of  radiation reaction force, several particles can reach higher energies. Part of the particles have $\gamma>250$, which is impossible if an electron moves along the attractor.\par
\begin{figure}[h]
	\centering
		\includegraphics[width=0.49\textwidth]{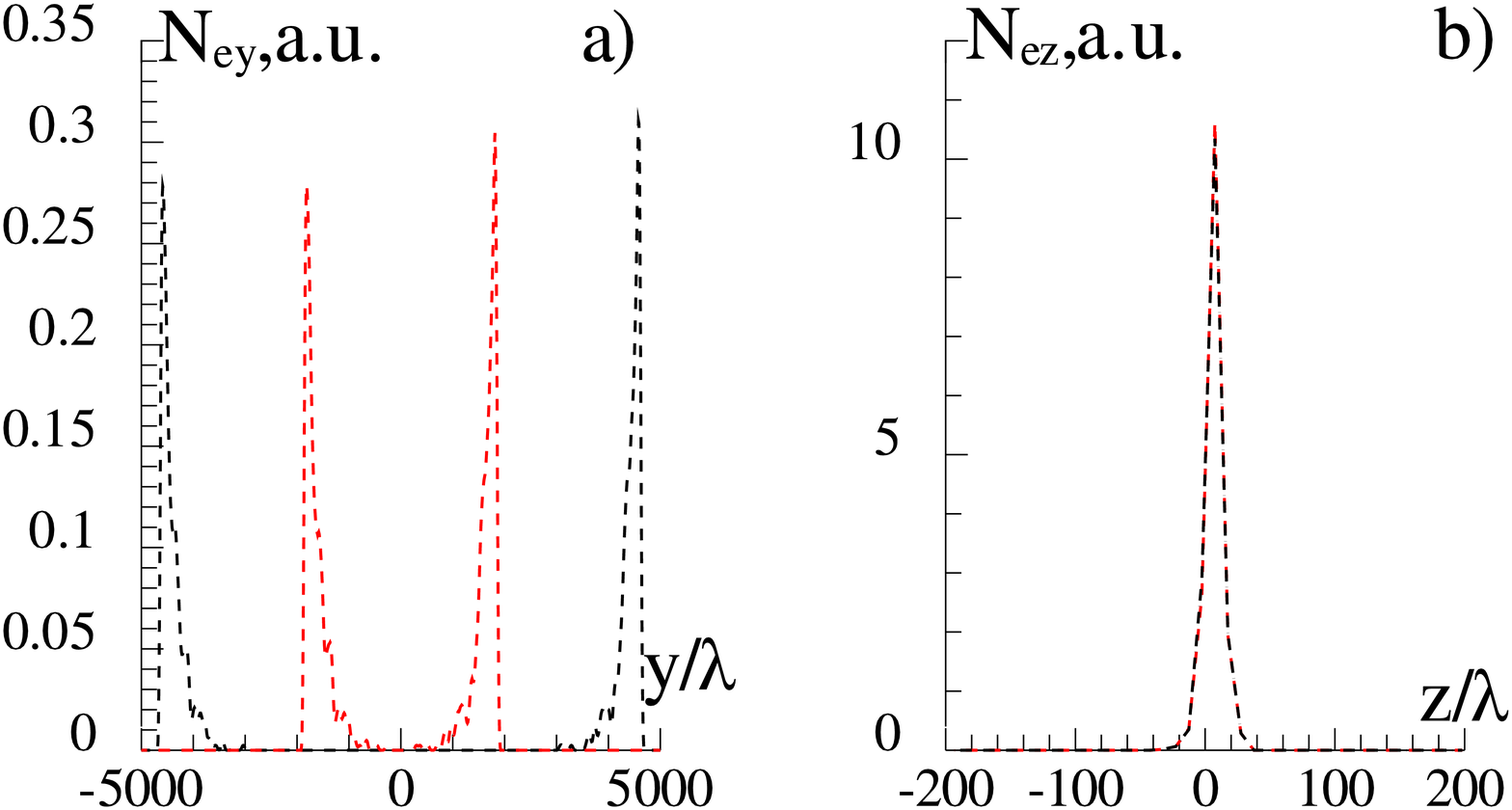}
	
	\caption{Electron distribution along Y and Z axes after 2000 (red dashed line) and 5000 (black dashed line) wave periods.}
	\label{fig:YZdr}
\end{figure}

\begin{figure}[h]
	\centering
		\includegraphics[width = 0.49\textwidth]{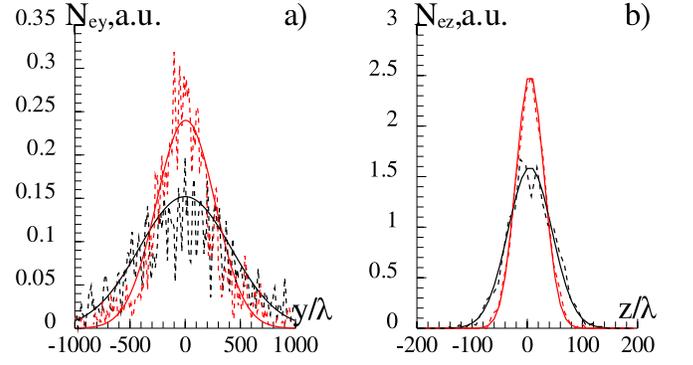}
	\caption{Diffusion of electrons along Y (a) and Z (b) axes. Black and red lines correspond to time t=2000T and t=5000T. Chain line corresponds to numerical calculation of 1000 trajectories of electrons initially distributed in the $\lambda/2$ range along Y and Z axes. Solid line corresponds to solution (\ref{sol_diff})}
	
	\label{YZdrwo}
\end{figure}

The electron energy spectrum reflects the specific features of particle motion. There are divergent electron beams along the characteristic curves $\upsilon t-y$ with radiation reaction. The electron distribution along the Z axis ($\sim20\lambda$) is determined only by the initial stage of approaching to the attractor in Fig. \ref{136YZt}a), \ref{fig:YZdr} and doesn't change after that. Thus, the electron temperature tends to zero. Note that in $\sim1000T$, the particles escape from the area where they were initially located. Figure \ref{fig:YZdr}a) confirms that the position of the leading edge of the electrons corresponds to $0.92ct$. 
If the radiative losses are neglected, the diffusive spreading of the electron distribution along the Y and Z axes is observed due to particle trapping by the antinode of the magnetic field and changing direction of motion (Fig.\ref{136YZt}b), \ref{YZdrwo}). In this case, the particle motion can be described by the diffusion equation for the linear density of the particles:
\begin{equation}
\frac{\partial N_e}{\partial t}=D\frac{\partial^2 N_e}{\partial r^2},
\label{diff_eq}
\end{equation}
where $r$ is implied to be $y$ or $z$. As can be seen from Fig. \ref{136YZt}, the diffusion along the Y axis is much faster, so it is possible to consider the diffusion in each direction independently. Since the electrons are initially within the half wavelength range and their following motion is considered during 5000 periods of the standing wave, during this time a typical scale of electron distribution becomes $\gg \lambda$. Then, it is possible to use the fundamental solution, assuming the diffusion coefficient to be constant and the initial distribution to be a delta function multiplied by the number of particles (\ref{sol_diff}):
\begin{equation}
N_{er}=\frac{N_0}{\sqrt{4\pi D_{\|,\bot}t}}\exp\left(-\frac{r^2}{4D_{\|,\bot}t}\right),
\label{sol_diff}
\end{equation}
where $N_{ey,z}$ is the linear electron density along the Y,Z axes. The evolution of the electron density profile along the Y and Z axes is shown in Fig. \ref{YZdrwo}. 
Calculations show that, if the wave amplitude $a = 136$, then the longitudinal diffusion coefficient $D_{\|}\sim120$, and the transverse one $D_{\bot}\sim1$.\par
Comparison of the electron motion with and without radiation reaction points that the qualitative differences in the motion occur only after the particles approach the attractor, and it takes hundreds of wave periods and continuous action of dissipative force in time. However, an electron emits a portion of its energy, quanta, and each act of emission "`knocks"'(or pushes) an electron, thus leading to lengthening of the time of convergence to the attractor or disappearance of the attractor at all. To understand the influence of the discreteness of photon emission on electron motion, we can estimate how often the electron emits and what portion of its energy turns into radiation. For this purpose it is necessary to estimate the parameter $\chi$. If the electromagnetic field is specified by (\ref{ef},\ref{bf}), $\chi$ can be written as
\begin{multline}
\chi=\frac{\hbar\omega}{mc^2}a\sqrt{\left(\rho_y^2+1\right)\cos^2(\tau+\varphi_0)\cos^2(\eta)}\\\overline{+\rho^2\sin^2(\tau+\varphi_0)\sin^2(\eta)}\\\overline{-0.5\rho_y\gamma\sin(2\eta)\sin(2(\tau+\varphi_0))}
\label{chi_expr}
\end{multline}
Rough estimation is $\chi\sim\frac{\hbar\omega}{mc^2}a^2\sim0.05$, which demonstrates a good fit with the value of $\chi$ along the trajectory with and without classical radiation reaction (Fig.\ref{fig:traj0},\ref{fig:traj1}). In the ultrarelativistic case, an average photon energy can be estimated as  $\epsilon_{ph}/mc^2\approx 0.45\gamma\chi\approx a^3\hbar\omega/mc^2\sim3.3$ \cite{Landau2}, and the classic radiation power $P_{rad}=\frac{2}{3}\frac{e^2}{\hbar c}\frac{mc^2}{\hbar\omega}\chi^2mc^2$ holds valid in the case $\chi<<1$. Hence, the average number of photons during the period can be assessed to be $N_{ph}\approx\frac{8.8\pi}{3}\frac{e^2}{\hbar c}\frac{mc^2}{\hbar\omega}\frac{\chi}{\gamma}\approx\frac{8.8\pi}{3}\frac{e^2}{\hbar c}a\approx9$, and part of electron energy lost in one act of photon emission is $\chi/2$. Note that, as follows from calculations, each of 1000 electrons in the wave field with amplitude $a=136$ during 5000 field periods emits $N_{ph}=6$ on the average. Rough estimations are in a good agreement with numerical results. Let us determine an average number of photons emitted by an electron over the wave period according to the probability of photon emission per unit time in the quantum case \cite{BayerKatkov}:

\begin{equation}
W=\frac{1}{3\sqrt{3}\pi}\frac{e^2}{\hbar c}\frac{mc^2}{\hbar \gamma}\int\limits_0^\infty \frac{5u^2+7u+5}{(1+u)^3}K_{2/3}(2u/3\chi)du.
\label{quant_prob}
\end{equation}

When $\chi\ll1$, the expression (\ref{quant_prob}) reduces to $W=\frac{5}{2\sqrt{3}}\frac{e^2}{\hbar c}\frac{mc^2\chi}{\hbar \gamma}$. This gives an estimate of a photon number $N_{ph}=\frac{5\pi}{\sqrt{3}}\frac{e^2}{\hbar c}a\approx9$, and each act of photon emission takes away about 0.02 part of electron's energy. Thus, $(2\pi\delta a^3)^{-1}\sim5$ wave periods are sufficient for an electron to lose a steady component of the momentum along the electric field. It may seem that the classical description can correctly describe electron motion due to a small fraction of electron energy converted to photon energy, but discreteness of emission plays an important role. On the one hand, the energy loss must lead to formation of an attractor. On the other hand, randomness leads to divergence of the particles in phase space. So, existence of an attractor is determined by the competition of these two factors. Estimations show that the particles move about 0.17 wave periods between the acts of photon emission without radiation reaction, and in this case the maximal Lyapunov exponent is $\lambda_1=0.2$. This means that the particles which are close in phase space diverge in $\sim T/(2\pi\lambda_1)\sim0.8T$. The time of particle convergence due to the radiation reaction may be estimated to be $\sim T/(2\pi\lambda_2)\sim24T$. The difference of these times destroys the attractor, so it is necessary to consider radiation as a quantum mechanism.\par
The importance of a quantum mechanism may also have a different explanation. 
Suppose that there is a regular attractor with allowance for quantum description of photon emission. Then, the attractor will be like in the classical case neglecting discreteness of radiation, since if $\chi\ll1$, then radiation losses are the same in the classical and quantum cases. Hence, the value of the characteristic Lyapunov exponents must be the same too. In this case, as was shown earlier, the largest non-zero Lyapunov exponent is $|\lambda_{mRR}|\ll0.1$, which corresponds to the basis unit vector (\ref{rb}). Perturbation along this vector during the time period between the radiation events $\Delta\tau_{rad}$ decreases as $e^{-|\lambda_{mRR}|\Delta\tau_{rad}}$. However, taking into account the discrete nature of radiation, it follows that the particle during $\Delta\tau_{rad}$ is affected only by the Lorentz force, and in this case the largest non-zero Lyapunov exponent $\lambda_{md}\approx0.2\gg|\lambda_{mRR}|$, and the perturbations grow along the vector (\ref{rb}) as $e^{\lambda_{md}\Delta\tau_{rad}}$. The result is that the perturbations grow as $e^{(\lambda_{md}+\lambda_{mRR})\Delta\tau}$, since $\lambda_{md}+\lambda_{mRR}>0$, and a regular attractor cannot exist. Consequently, the discreteness of radiation plays a qualitatively important role. Note that the system of equations (\ref{llpy}-\ref{llz}) when $a\gg1$ is determined by one similarity parameter $\delta a^3\propto \omega a^3$. Thus, in contrast to the classical case, there is no regular attractor and suppression of stochastic heating, if $(0.0006/\delta)^{1/3}\le a\le(0.037/\delta)^{1/3}$ and $\omega\ll0.05c/r_e$ or $\lambda\gg100 r_e\approx0.001 A$.

\section{Electron motion taking into account discreteness of radiation}

Let us use the tried-and-true approach for the quantum description of radiation and classical description of electron motion \cite{Bell_Kirk_Monte_Carlo}. Electron motion is described by equations \ref{landlifp},\ref{landlifr}, but without radiation reaction. Each particle should propagate a randomly assigned optical length in the electromagnetic field before photon emission. If the random number $r_1\in(0,1)$, then the optical path $L_{opt}=\log{\frac{1}{1-r_1}}$. Once $\int Wdt=L_{opt}$, then the procedure of photon emission is started. The choice of the photon energy is carried out with the help of the Monte Carlo Rejection sampling method \cite{Neal_AS}. To do so, the auxiliary function $M(\eta)$ is selected, such that $M(\eta)\cdot C(\chi)\ge dW(\chi,\eta)/d\eta$ where $\eta=\epsilon_{ph}/(mc^2\gamma)$ is the fraction of the electron energy converted to the photon energy:
\begin{equation}
\label{M_eta}
M(\eta)=\begin{cases}\frac{0.325}{\eta^{2/3}},& \eta\le0.87 \\ \frac{0.18}{(1-\eta)^{1/3}}, & 0.87<\eta\le1\end{cases}
\end{equation}

\begin{equation}
\label{C_chi}
C(\chi)=1.6\frac{e^2}{\hbar c}\frac{mc^2}{\hbar\omega\gamma}\chi^{2/3}.
\end{equation}
Here, $\chi$ is calculated according to the momentum and strength of magnetic and electric fields in the constant-field approximation; $dW(\chi,\eta)/d\eta$ is defined as follows \cite{BayerKatkov}:
\begin{gather}
\frac{dW(\chi,\eta)}{d\eta}=\frac{1}{\pi\sqrt{3}}\frac{e^2}{\hbar c}\frac{mc^2}{\hbar\omega\gamma}\left[\int\limits_{\frac{2\eta}{3(1-\eta)\chi}}^\infty K_{5/3}(y)dy+\right.\nonumber\\
\left. \frac{\eta^2}{1-\eta}K_{2/3}(\frac{2\eta}{3(1-\eta)\chi}) \right].
\label{dWdeta}
\end{gather}
The second random number $r_2\in(0,1)$ allows determining $\eta(\chi,r_2)$ from the equation $M(\eta)=r_2$. Using the third random number $r_3\in(0,1)$ it is possible to determine whether the photon is emitted $r_3<dW(\chi,\eta)/d\eta/(C(\chi)*M(\eta))$; otherwise, it is necessary to choose a random number $r_2$ again. A properly chosen function $M(\eta)$ is as close as possible to the function $dW(\chi,\eta)/d\eta/C(\chi)$, which reduces the number of attempts to emit a photon. Also, it is easy to determine $\eta(\chi,r_2)$, i.e. to find the inverse function.\par
\begin{figure}[h]
	\centering
		\includegraphics[width=0.49\textwidth]{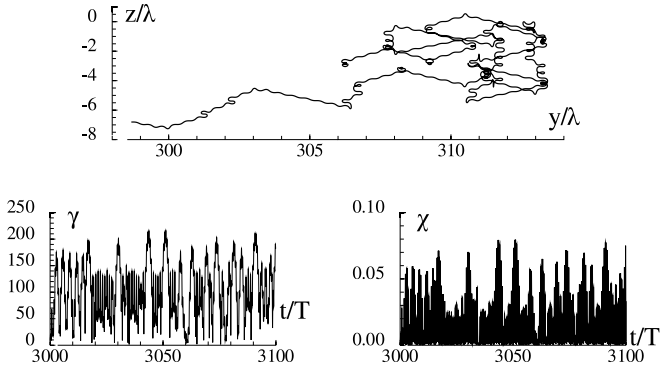}

	\caption{Electron trajectory as in Fig.\ref{fig:traj0} but with quantum description of photon emission.}
	\label{fig:trajq}
\end{figure}

Let's consider a typical trajectory of an electron using the quantum description of radiation (Fig.\ref{fig:trajq}). 
Trajectories in cases of quantum description and without radiation losses are similar, as is evident from comparison of Figs. \ref{fig:trajq}, \ref{fig:traj0}, but there is one qualitative difference. Without radiation reaction, there is an integral of motion $p_z-A(y,t)=\bf{const}$, because the equation of motion does not depend on the coordinate $z$. Therefore, the motion along the z coordinate is different for different initial phases of a standing wave, the electrons can have a steady component of momentum along the Z axis, but this component vanishes both with allowance for classical and quantum radiation losses.\par
\begin{figure}[h]
	\centering
		\includegraphics[width = 0.23\textwidth, height =0.15\textwidth]{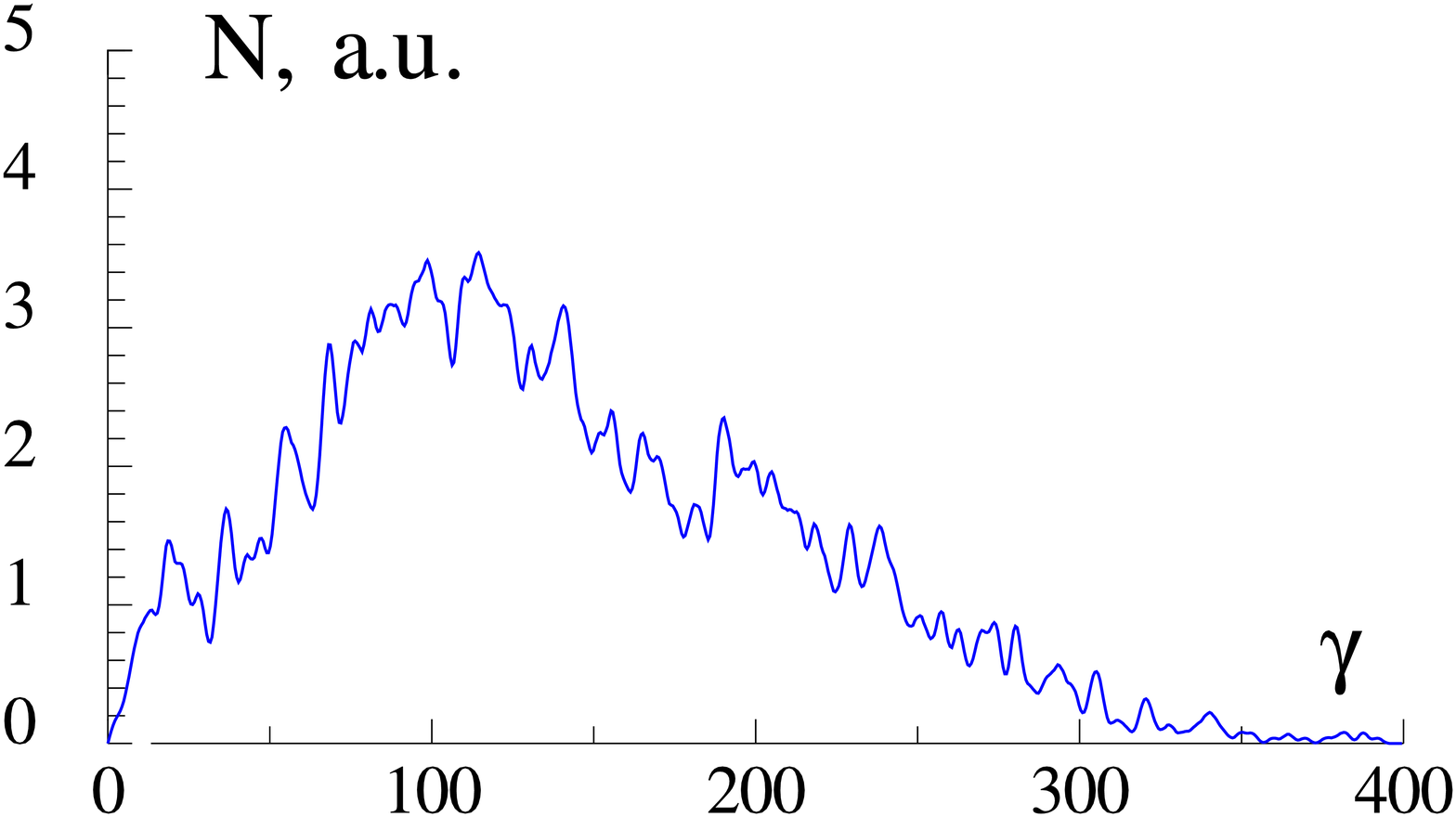}
		\includegraphics[width = 0.23\textwidth, height = 0.15\textwidth]{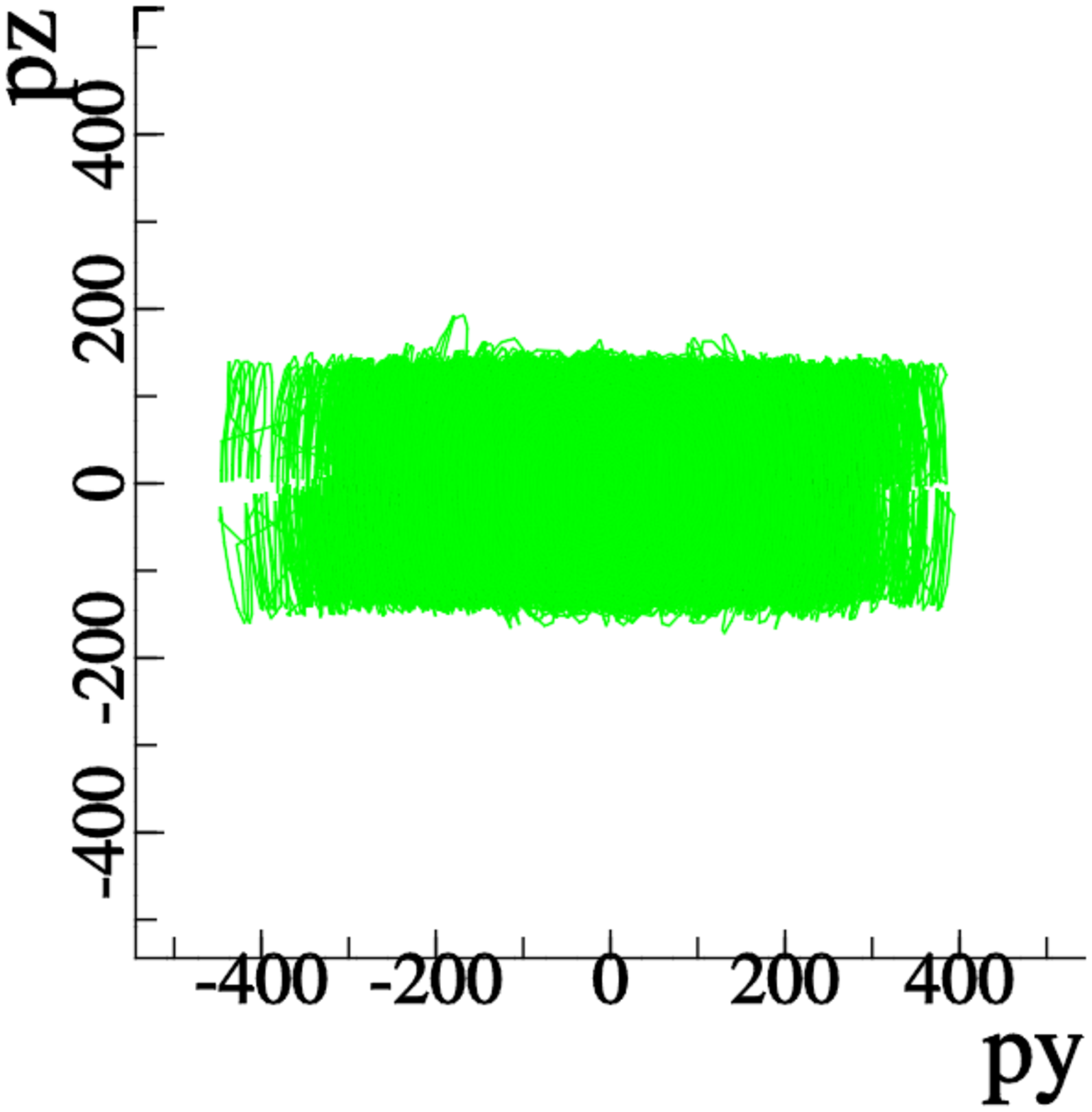}	
	\caption{Electron energy distribution in the field of a plane wave with amplitude $a=136$ (a), and phase space projection on the $p_yp_z$ plane (b) with quantum description of photon emission.}

	\label{fig_af_65_50Enq}
\end{figure}
\begin{figure}[h]
	\centering
		\includegraphics[width = 0.45\textwidth, height=0.2\textwidth]{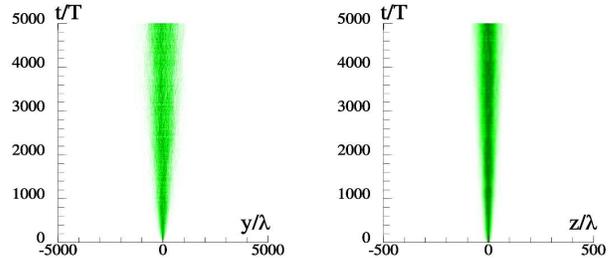}
	\caption{Electron distribution along Y and Z axes as a function of time for $a=136$ using quantum description of photon emission.}
	\label{136YZq}
\end{figure}
\begin{figure}[h]
	\centering
		\includegraphics[width = 0.49\textwidth]{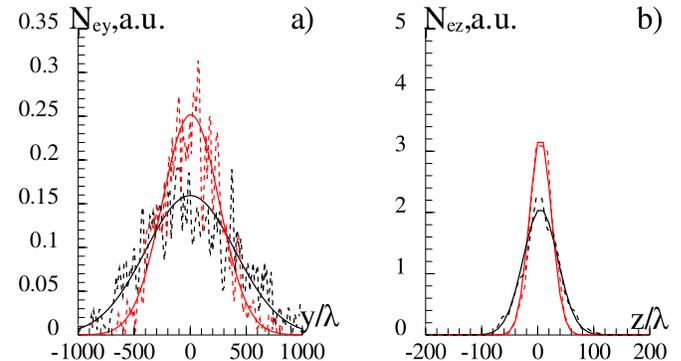}
	\caption{Diffusion of electrons along Y (a) and Z (b) axes at quantum description of photon emission. Black and red lines correspond to time t=2000T and t=5000T. Chain line corresponds to numerical calculation of 1000 trajectories of electrons distributed initially in $\lambda$ range along Y and Z axes. Solid line corresponds to solution (\ref{sol_diff})}
	
	\label{YZdrq}
\end{figure}

Let us now examine electron bunch motion under the initial conditions same as in the classical case. Qualitatively, the motion of the electron bunch coincides with the case without radiation friction, which confirms the conclusion of the importance of discreteness and that the motion is generally determined by the Lorentz force rather than the radiation reaction (Fig.\ref{fig_af_65_50Enq},\ref{136YZq}). Figure \ref{136YZq} shows that part of particles may have energy greater than $250mc^2$. The spectrum has an analogous feature, when the radiation reaction is not taken into consideration. This is due to the discrete nature of radiation, because particle does not undergo radiation damping between the acts of emission. The electron bunch motion can also be described using the fundamental solution \ref{sol_diff} (Fig.\ref{YZdrq}). Calculations demonstrate that the diffusion coefficients are smaller than in the case without radiation friction, as the electron changes its direction more often. Thus, the diffusion decreases and the electron bunch "spreads" slower. 

The longitudinal diffusion coefficient is $D_{\|}\sim100$, and the transversal diffusion coefficient $D_{\bot}\sim0.6$. In order to determine why the diffusion is like this, note that the motion of an electron in a plane wave with quantum description of radiative losses is chaotic. It is found that the electron retains the character of the motion each half period, and the character of motion in the next half period is determined by the present one. Such motion can be regarded to be a Markov chain.

\section{Markov chain}

\begin{figure}[h]	\centering
		\includegraphics[width = 0.4\textwidth]{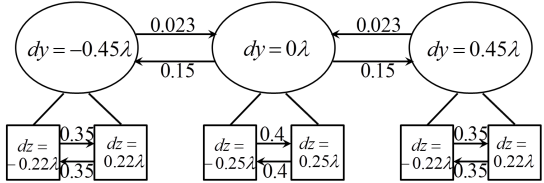}
	\caption{Schematic electron trajectory as Markov chain}
	\label{SchemeTr}
\end{figure}

\begin{figure}
	\centering
		\includegraphics[width=0.49\textwidth]{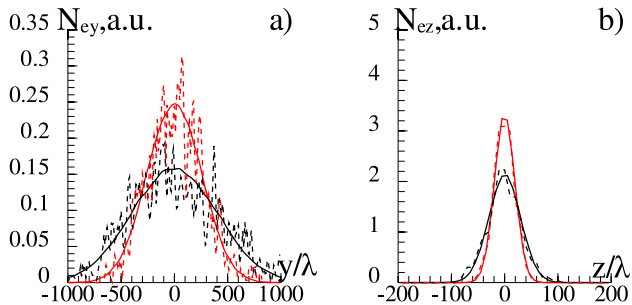}
	\caption{Electron distribution over 2000 wave periods (red line) and over 5000 wave periods (black line) along Y a) and Z b) axes as a result of Markov chain process(solid line) and as a result of numerical solution of motion equation (dashed line) taking into account quantum description of radiation.}
	\label{fig:YZrm}
\end{figure}

The process described by the Markov chain is defined by a transition matrix which sets the probabilities of the states of the n${}^{th}$ step from the known state of the n-1${}^{th}$ step. We use the formalism of Markov chains to determine the width of the electron distribution along the Y and Z axes. There are 3 states of motion along the Y axis (all the states are shown in Fig. \ref{SchemeTr}). 1) A particle shifts by $0.45\lambda$ in the direction opposite to the Y axis, while shifting by $\pm0.22\lambda$ along the Z axis. The probability of changing the direction with respect to Z without changing the motion along Y is 0.35. The probability of the transition to the second state is 0.023. 2) On the average, the particle propagating $\pm0.25\lambda$ along Z does not move along Y. The probability of changing the direction of motion along the Z axis is 0.4. The probability of transition to state 3) or 1) is 0.15. State 3) is the same as state 1), but the electron moves along the Y axis. Numerical simulations allow determining the probability of the transition between the states (indicated above the arrows in Fig. \ref{SchemeTr}), as well as determining the states.
Assume that $S^n=\left(\begin{smallmatrix}S_1^n\\S_2^n\\S_3^n\end{smallmatrix}\right)$ is the state vector of electron motion along the Y axis corresponding to the n${}^{th}$ step. $S_1$ and $S_3$ denote the probability of the motion along the Y axis and in the opposite direction with respect to the Y axis correspondingly, $S_2$ denotes the probability of rest along the Y axis. Then the transition matrix $P_{ij}$ from state i to j is written in the form 
\begin{equation}
P_{ij}=\begin{pmatrix}
0.977 & 0.023 & 0\\
0.15& 0.7 & 0.15\\
0 & 0.023 & 0.977
\end{pmatrix} 
\label{Pmy}
\end{equation}
which means that $S^n$=$\left(P^T\right)^n S^0$. 
Let's calculate the mathematical expectation $\mu$ and the variance $D$ of electron deviation along the Y axis in n steps $\Delta Y_\Sigma=\sum\limits_{i=1}^nY_i=\sum\limits_{i=1}^n (-0.45\lambda,0\lambda,0.45\lambda)S^i$. The mean value of the sum is the sum of the mean values, i.e. $\mu\delta Y_\Sigma=\sum\limits_{i=1}^n \mu (-0.45\lambda,0\lambda,0.45\lambda)S^i$. Initially, the particles are distributed uniformly in states 1), 2), and 3) according to numerical simulations of electron bunch dynamics, i.e. $S^0=\left(\begin{smallmatrix}1/3\\1/3\\1/3\end{smallmatrix}\right)$. $\mu\delta Y_\Sigma=0$ because of the equivalent motion in both directions along the Y axis and because of the symmetry of matrix $\left(P^T\right)^n$ and the vector of initial states. Thus, as evidenced by Fig. \ref{YZdrq}, on the average, motion occurs without drift of the center of mass. Let's calculate the variance $D\Delta Y_\Sigma=\sqrt{\sum\limits_{i=1}^n\sum\limits_{j=1}^n \mu(Y_i Y_j)}=\sqrt{\sum\limits_{i=1}^n\mu Y_i^2+2\sum\limits_{i=1}^n\sum\limits_{j>i}^n\mu(Y_iY_j)}$. 
Note that the events become weakly correlated after a sufficiently large number of steps and the state vector at the $k$-th step $(P^T)^kS^0=S_{av}$. In order to find $S_{av}$, it is necessary to solve the equation $P^TP_{av}^T=P_{av}^T$ in conjunction with the condition that the sum of the values in the rows of equilibrium transition matrix $P_{av}$ (a matrix with three identical columns) is equal to 1. Because of the symmetry of motion in both directions along the Y axis, the equilibrium transition matrix has the form $P_{av}=\begin{pmatrix} a& b & a\\a& b & a\\
a & b & a\end{pmatrix}$, $2a + b = 1$, and $0.977a + 0.15b = a$. Hence, $a = 0.4644$, $b = 0.0712$, and $S_{av}=\left(\begin{smallmatrix}a\\b\\a\end{smallmatrix}\right)$. It took $N_{eq}\approx500$ steps for the equilibrium state to be formed. The electron bunch dynamics is observed during 5000 wave periods, i.e. $N_{st}=10000$ steps. The first term in the expression of variance can be calculated as $\sum\limits_{i=1}^n\mu Y_i^2=n\mu Y_{av}^2=0.188n\lambda^2$, where $Y_i$ is the displacement during half  the wave period along the Y axis. Let's calculate the second term. Assume that $S^i$ is the state vector at the i$^{th}$ step. For finding the correlation of states i and j, it is necessary to know the state vector $S^j$ using the conditional probability, if we know the value of random variable $Y_i$. These probabilities are expressed in the form of elements of the matrix $(P_{kl})^{j-i}$, where the $l^{th}$ column indicates which states will be at the j$^{th}$ step. If the state k is at the i$^{th}$ step, then $\sum\limits_{i=1}^n\sum\limits_{j>i}^n\mu(Y_iY_j)=\sum\limits_{i=1}^n\sum\limits_{j>i}^n \left(\begin{smallmatrix}-0.45\lambda \\0\lambda \\0.45\lambda \end{smallmatrix}\right)^T\left(P^T\right)^{j-i}\left(\begin{smallmatrix}-0.45\lambda&0&0\\0&0&0\\0&0&0.45\lambda\end{smallmatrix}\right)\left(P^T\right)^iS^0$. The approximation $(P^T)^kS^0=S_{av}$ is used because $N_{st}\gg N_{eq}$, hence $A=\left(\begin{smallmatrix}-0.45\lambda&0&0\\0&0&0\\0&0&0.45\lambda\end{smallmatrix}\right)\left(P^T\right)^iS^0=\left(\begin{smallmatrix}-0.21\lambda\\0\\0.21\lambda\end{smallmatrix}\right)$.
Note that $(P^T)^iA$ tends to zero for large i, since components 1 and 3 of the vector A are similar in magnitude and opposite in sign, in other words the dependence on initial conditions disappears and the motions along the Y axis are equal in both directions. Thus, for good accuracy it is sufficient to take into account a certain quantity of terms $(P^T)^iA$. It suffices to take into account only 500 terms in view of the fact that the equilibrium state is achieved in 500 steps. Finally, to a good accuracy $\sum\limits_{i=1}^n\sum\limits_{j>i}^n\mu(Y_iY_j)=n\sum\limits_{j=1}^{500}\left(\begin{smallmatrix}-0.45\lambda \\0\lambda \\0.45\lambda \end{smallmatrix}\right)^T(P^T)^iA=7.9n\lambda^2$. Generally, the variance is due to correlations and $D\Delta Y_\Sigma=4\sqrt{n}\lambda$. Based on the root function of the number of steps (twice the number of periods) it becomes obvious that "spreading" of an electron bunch occurs at large n with constant dimensionless diffusion coefficient $D_{\|}=4^2\cdot2\pi\approx100$. The value of the variance is in a good agreement with the characteristic width of the electron distribution along the Y axis in Fig. \ref{YZdrq} and with the diffusion coefficient. Figure \ref{fig:YZrm} shows the distribution of electrons along the Y and Z axes derived from the Markov process according to Fig. \ref{SchemeTr} in comparison with the numerical simulation of electron motion taking into account the quantum description of radiation.\par

Similar analysis can be made with the electron motion along the Z axis. In this case, the state vector consists of four elements. The result is the following: $D\Delta Z_\Sigma=0.3\sqrt{n}\lambda$ and $D_{\bot}=0.3^2\cdot 2\pi\approx0.6$, which agrees well with the results presented in Fig. \ref{YZdrq}.\par
It should be noted that the consideration of the electron motion as a process described by a Markov chain allows us to understand the transition to the normal radiative trapping \cite{Lehmann_PRE},\cite{ART_Gonoskov}. The state of $Y_i=0\lambda$ becomes absorbing, i.e. in terms of the transition matrix we can say that $P_{22}\rightarrow1$, $P_{21}$ and $P_{23}$ tend to zero with increasing $a$.

\section{Conclusion}

The electron motion in a plane standing wave is considered within the framework the quantum and classical description of the radiation reaction. It is shown that in the case of small $\chi$, the classical description of photon emission can lead to qualitatively wrong electron dynamics despite the small radiative losses. The Lyapunov characteristic exponents obtained in this case show that there is a regular attractor in the phase space, whereas without radiative losses the motion is chaotic. In the quantum case, the character of motion changes drastically due to the discrete nature of photon emission, which randomizes particle motion. This leads to the additional divergence in the phase space, which completely destroys formation of a regular attractor. A new method is proposed for describing electron kinetics in this case using Markov chains. The electron distribution obtained by this method is in a good agreement with the results of direct calculations.\par
The research is partly supported by the grant (the agreement of August 27, 2013 №02.В.49.21.0003 between The Ministry of Education and Science of the Russian Federation and Lobachevsky State University of NizhniNovgorod) and RFBR (grant №14-02-31495 mol$\_$a).

\bibliography{Bibl1}
\end{document}